# TRIFECTA: Security, Energy-Efficiency, and Communication Capacity Comparison for Wireless IoT Devices

Shreyas Sen, Jinkyu Koo, Saurabh Bagchi[Ψ]

## ABSTRACT
The widespread proliferation of sensor nodes in the era of Internet of Things (IoT) coupled with increasing sensor fidelity and data-acquisition modality is expected to generate 3+ Exabytes of data per day by 2018. Since most of these IoT devices will be wirelessly connected at the last few feet, wireless communication is an integral part of the future IoT scenario. The ever-shrinking size of unit computation (Moore's Law) and continued improvements in efficient communication (Shannon's Law) is expected to harness the true potential of the IoT revolution and produce dramatic societal impact. However, reducing size of IoT nodes and lack of significant improvement in energy-storage density leads to reducing energy-availability. Moreover, smaller size and energy means less resources available for securing IoT nodes, making the energy-sparse low-cost leaf nodes of the network as prime targets for attackers. In this paper, we survey six prominent wireless technologies with respect to the three dimensions—security, energy efficiency, and communication capacity. We point out the state-of-the-art, open issues, and the road ahead for promising research directions.

## Keywords
Internet of Things (IoT), Wireless, Communication, Energy-Efficiency, Security

## 1. INTRODUCTION

**Cheap Ubiquitous Computing → Smart Things:** Through five decades of continued scaling, following Moore's Law, the size of unit computing has gone to virtually zero. Starting with mainframe computers in the 60's, that used to be of the size of a room, we saw continuous reduction in the size of a computer. We saw computers progress through the Mini, the Workstation, the Personal Computer (PC) down to Laptops in the 2000's (Figure 1). The decade of 2010 was dominated by Mobile devices (e.g. Smartphones). By the year 2020, the size of unit (meaningful) computation will be so small that it will be barely visible. This will enable cheap, ubiquitous computation all around us, incorporated into everyday things like wearables, household devices, and mobile payment devices. The ability to incorporate significant computation in an almost invisible footprint is transforming everyday objects into *Smart Things*.

**Cheap Wireless Connectivity → Connected Things:** The emergence of the Internet as a household commodity worldwide coupled with tremendous progress in commoditization of wireless connectivity (especially cellular 5G and wireless LAN) means that billions of things can now be wirelessly connected to the Internet.

**Smart Connected Things → Internet of Things (IoT):** The emergence of cheap computing following Moore's Law is enabling Smart Things and emergence of cheap wireless connectivity following Shannon's law (Figure 1) is creating smart connected things. At present, we are standing at the crossroads of smart and connected internet of things (IoT), which is quickly transforming human lives. The number of internet connected devices has already passed the number of human beings on the planet in 2009 and is increasing exponentially. Cisco estimates that by 2020, there will be 3.4 devices and connections per person.

**Smaller Size, Similar Energy-Density → Lower Available Energy**: Though the size of unit computation is falling fast, the energy-storage/battery technology is improving only very slowly, leading to a reducing amount of available energy in smaller nodes. Due to its small footprint, the size of the battery included in such sensor nodes is limited. Moreover, including a battery means increased deployment cost and more importantly maintenance cost (to change the battery periodically). Since the electronics lifetime

The authors are with Purdue University, School of Electrical and Computer Engineering, e-mail: {shreyas, kooj, sbagchi}@purdue.edu.
Ψ Contact author: Saurabh Bagchi

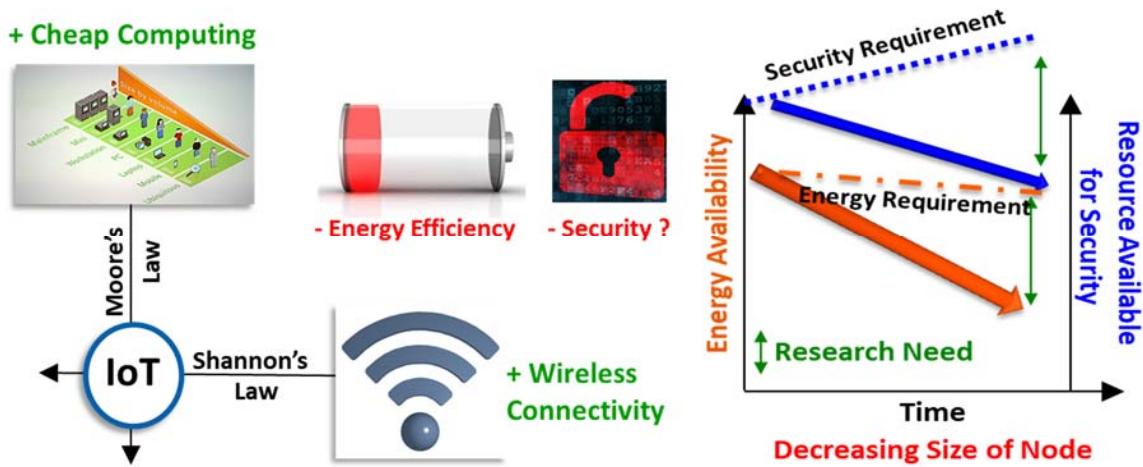

Figure 1: IoT as the junction of Moore's Law and Shannon's Law. A big bottleneck for Ubiquitous Computing using IoT Nodes is reduced energy availability and increased security vulnerabilities in leaf nodes as the size of the node decreases

is generally significantly higher than battery-lifetime, it is desirable to develop *Net-Zero Energy* sensor nodes, that perpetually run on harvested energy. Whereas the trend is to pack more and more functionality even in energy-constrained nodes and they need to communicate wirelessly. This leads to energy-gap and calls for significant improvement in energy-efficiency for computing and communication in energy-constrained nodes.

**Smaller Size, Lower-Energy → Lower Resource Availability for Security:** It is well-known that the security of a network is often only as good as its weakest link. Energy-sparse, size-constrained IoT nodes have limited resources to guarantee strong security and hence often are the weakest link in the end-to-end system. While the resource available for security is reducing (Figure 1, right) with reducing size, the security requirements of these leaf nodes are increasing, creating a strong need for research in lightweight, resource-constrained security technologies.

In many of the compelling application areas, the security of the communication channel is of primary importance, including the possibility of eavesdropping (i.e., loss of confidentiality) and denial of service (i.e., loss of availability) [1]. The two concerns that have traditionally been looked at for this class of systems are energy-efficiency, and communication capacity [2]. In this paper, we analyze prominent wireless technologies for IoT with respect to the three dimensions—security, energy efficiency, and communication capacity. These dimensions are of course inter-related, e.g., an otherwise energy-efficient system may become unusable if it needs cryptographic protocols and that is expensive on such systems.

## 2. PROMINENT TECHNOLOGIES

Multiple classes of wireless technologies, namely, Wireless Local Area Network (WLAN: Wi-Fi, Bluetooth), Sensor Network (ZigBee), Near-field (NFC and emerging High-speed Proximity Communication [3]), and wide-area wireless communication (LoRa), will be compared across security, energy efficiency, and communication capacity.

### 2.1 Energy-Efficiency and Communication-Capacity

**State-of-the-art:** Table 1 summarizes several wireless techniques in terms of communication distance supported, typical energy-efficiency (in Joules/bit), data rate (communication capacity in Bits per second) and security. Figure 2 plots these technologies on the energy-efficiency (Y-axis) vs. Data rate (X-axis). It is to be noted, that the maximum data rate are often limited by FCC and the standard. The communication energy-efficiency varies from several from pJ/b to µJ/b, *i.e.*, six-orders of magnitude, depending on the PHY and the distance supported. Significant amount of this energy is wasted due to inflexible, worst-case radio designs [4].

**Energy-Gap:** Current battery technology supports enough energy for low-performance communication, and hence we are seeing a plethora of commercial low-performance battery-operated IoT devices. However, mobile devices are severely energy-constrained for both battery-operated high-performance devices and energy-harvested low-performance devices. A typical smartphone battery holds 5-10 Watt-hour of energy. Communicating 10 Gbps data (*e.g.*, 4K video, 30fps, RGB, 12b color depth = 9.56 Gbps raw) at 1nJ/b means 10 Watt power. Hence the mobile battery runs out within an hour, just supporting such communication, let alone processing and display. Similarly, for energy-harvested devices, solar harvesting lends tens of mWs of power in favorable outdoor conditions. However, for all other modalities (*e.g.*, indoor lighting, vibration, thermal, and RF harvesting) typical harvested power falls in the range of 50-200µW. For a sensor node trying to communicate 1Mbps (*e.g.*, compressed, intermittent video) with energy efficiency of 1nJ/b, it will consume 1mW just for the communication portion. This highlights the energy-gap present for current IoT sensor nodes. An order of magnitude improvement in communication energy-efficiency will open up many applications of ubiquitous connected IoT nodes.

### 2.2 Security considerations

As we gradually move toward using some of the smart devices for critical operations, security will become a primary driver for which devices win out in the marketplace. We are already seeing some such uses around us, such as, in mobile payment systems (Google Wallet, Apple Pay, Samsung Pay) and wearable healthcare devices which monitor for critical health signals (such as, heart rate, VO2 level) and in case of critical indicators, communicate to a health provider. We survey here some of the successes and challenges for securing the wireless technologies under discussion here. We also discuss some of the unique aspects of security in this domain. In all of this, it is important to keep in mind that security should be considered as improving the state of affairs in one of the three axes (Figure 2) – confidentiality (of the information being stored or exchanged), integrity (of the data being stored), and availability (being able to access the device and its stored state). Also, the security achieved is hardly ever zero-or-one for any of these axes, but rather on a sliding scale.

**Geographical Proximity as an Aid**

Technologies that operate in very close proximity, such as NFC with < 20 cm range, rule out most man-in-the-middle (MITM) attacks. A typical MITM attack scenario is as follows, where

Table 1: Comparison of State-of-the-art Wireless Techniques for IoT nodes

|  | **Proximity** | **NFC** | **ZigBee** | **BT** | **WiFi** | **LoRa** |
|---|---|---|---|---|---|---|
| **Distance (m)** | 1mm | 10cm | 10-100m | 10-100m | 30-50m | ~km |
| **Data rate (bps)** | 8-32 G | 0.02-0.4 M | 0.02-0.2 M | 0.8-2.1 M | 300M (11g) 7Gbps (11ac,11d) | 200K |
| **Energy-Efficiency (J/b)** | 4p | 1n | 5n | 15n | 5n | 1u |
| **Security[1]** | H[1] | M | L[1] | L | M/H[1] | Relatively Unknown |

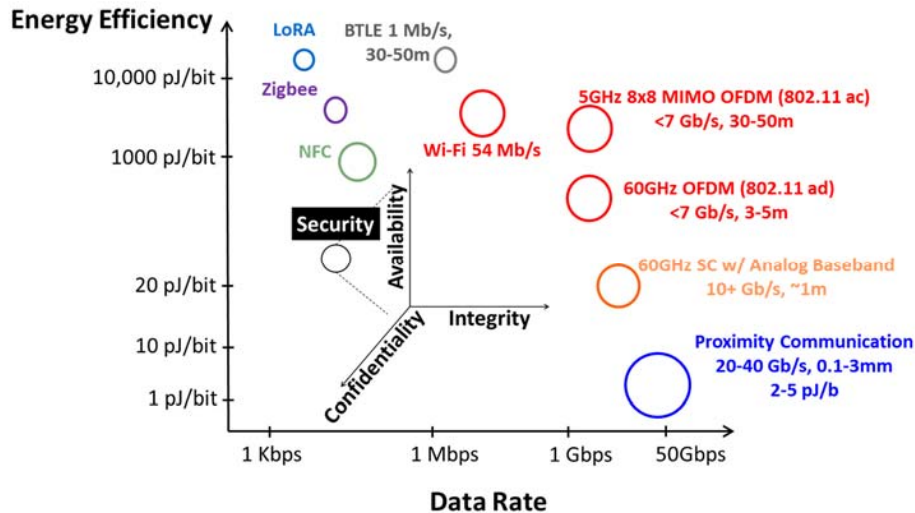

Figure 2: State-of-the-art energy-efficiency vs. Communication data rate prominent wireless IoT PHYs. The size of the circle represents the strengths of security, which consists of Confidentiality, Availability and Integrity. The gap in the bottom left motivates research need for low-speed, reliable yet highly efficient secure communication techniques

Mallory, an attacker, is interposing herself between the communication of two legitimate parties, Alice and Bob. This kind of MITM attack, uniquely to our proximal wireless communication scenarios, is possible even when cryptography is being used, due to the ability of the attacker to intercept the communication.

1. Alice sends her public key to Bob, but Mallory can intercept it. Mallory sends Bob her own public key for which she has the matching private key. Now Bob wrongly thinks that he has Alice's public key.

2. Bob sends his public key to Alice, but Mallory can intercept it. Mallory sends Alice her own public key for which she has the matching private key. Now Alice wrongly thinks that she has Bob's public key.

3. Alice sends Bob a message encrypted with Mallory's public key, but Mallory can intercept it. Mallory decrypts the message with her private key, keeps a copy of the message, re-encrypts the message with Bob's public key, and sends the message to Bob. Now Bob wrongly thinks that the message came directly from Alice and has no indication that the message has been intercepted and decrypted.

4. Similarly to step 3 above, when Bob sends Alice a message, Mallory can again decrypt it and optionally modify it, before passing it on to Alice pretending that it came from Bob.

This kind of MITM attack can be mitigated if Alice and Bob have visual connection due to geographical proximity and can prove to each other's devices that there is such proximity. This typically requires entering a secondary authentication token that appears on both devices, such as, a long random pin. It is in the choice of the secondary authentication mechanism that the capability of the device will become a crucial factor. For example, if the device has output display, then the pin can be displayed; if the device has a touch sensor, then Alice and Bob can be asked to authenticate by physically touching the other's device.

**Isolation and abstraction**

Isolation of different hardware and software modules has been considered a key building block for secure systems in the traditional desktop and server world. This means that there are boundaries to what each hardware or software module can access (e.g., only some parts of the device's memory) and thus if one module is compromised, the entire system does not get compromised. In the domain under consideration here, such isolation may or may not be possible depending on the specific device's price point. One example where such isolation is widely used today is in smart phones. In most smart phones, there is the relatively powerful main processor and a separate baseband processor [5]. The baseband processor runs the radio control functions, which have real-time requirements, and therefore a real-time OS runs on the baseband processor. However, due to the proprietary nature of the software stack on it, there are often security vulnerabilities found in these [6]. The software on the main processor trusts the software running on the baseband processor and thus the vulnerability can spread. Thus, we see that despite isolation, if the separation is not enforced, security breaches occur. Therefore, the correct design point is whenever isolation is possible, either in hardware or software, then enforcement of the separation boundary is needed. Some recent efforts with low-end embedded devices [7], [8] are showing how it is possible to enforce isolation with limited hardware support and mainly through software techniques. We acknowledge however that for many low-end smart things, such isolation will be infeasible and therefore systems must be built with an acknowledgment of the vulnerable nodes and understanding of their spread potential.

**Out-of-band mechanism for security**

An interesting interplay between multiple technologies happens in this space to provide increased security. Many security protocols rely on some out-of-band (OOB) mechanism for exchanging some critical information, which helps secure a communication channel. In authentication, OOB refers to utilizing two separate networks or channels, one of which is different from the primary network or channel, simultaneously used to communicate between two parties or devices for identifying a user. For example, a cellular network is commonly used for out-of-band authentication. An example of out-of-band authentication is when an online banking user is accessing her online bank account with a login and a one-time password (OTP) is sent to her mobile phone via SMS. The primary channel would be the online login screen where the user enters her login information and the OTP sent through the out-of-band channel.

In our domain, oftentimes there is a clear OOB which is the humans interacting through their respective devices [9]. This naturally allows certain levels of trust to be established among the communicating individuals. With the right security protocol, this trust can be transferred to devices that belong to the users, enabling two devices to establish a trusted communication channel that reflects the existing trust their users place on one another. A typical example is the pairing of two Bluetooth devices with active participation of the users. In the Bluetooth, Secure Simple Pairing (SSP) mode, it uses NFC for achieving security. One issue to keep in mind here is that the devices should be reasonably time synchronized, say to within 10s of milliseconds. Much more accurate time synchronization has been demonstrated even for ad-hoc wireless networks [1].

# 3. THE ROAD AHEAD

## 3.1 Communication Capacity and Energy-Efficiency

In this data-driven IoT revolution, workloads, operating conditions and computational/communication demands on distributed and connected devices will undergo large dynamic ranges of several orders of magnitude. Energy constrained IoT nodes will demand the highest possible energy efficiency across the entire range of operation under changing contexts. A context could be defined as channel conditions, applications, latency, QoS, data rate requirements, battery condition, process variation, among others. Current systems are typically over-designed to handle all possible context, which creates an unfavorable trade-off between fidelity and power efficiency. Learning from nature, we notice that a human brain, continuously adapts to its surroundings to perform more efficiently. It also self-learns [10] the optimum ways with experience. Similarly, in context-aware communication, a smart IoT device understands its own context and adapts itself "on-the fly" for optimal energy-efficiency and performance. Such context-aware communication could be divided in two distinct categories, namely Intra-PHY [11] and Inter-PHY adaptation, as described in [2]. In brief, the former means adapting within *one* physical layer wireless communication channel, while the latter involves multiple such channels.

Along with context-awareness, innovative technologies specific to emerging applications can enable order(s) of magnitude improvement in both communication capacity and energy-efficiency, even simultaneously. As an example, recently developed Capacitive Proximity communication [3] provides wire-like data rate (32Gbps) and energy-efficiency (4pJ/b) without a physical wire, and enables >100× benefit over short-range mm-wave communication, allowing high-speed transfer (*e.g.*, fast video, photo download from smartphone to laptop just by placing it on top of the laptop, without connecting a wire). Another example is Human Body Communication [12], that utilizes the conductive properties of human body to connect wearables and implantables, reducing Body Area Network (BAN) connectivity energy by >100x, while improving privacy, as the signals are mostly contained within the body and cannot be snooped from far away by an adversary. Similar application-specific technology developments will be needed to unlock energy efficiency of >10x and more.

Most importantly, since communication energy is often the bottleneck, it's wise to communicate 'information' than 'raw data' to and from the sensor nodes. This only makes sense if the energy cost of in-sensor information-extraction (*i.e.*, in-sensor analytics) is significantly lower than the communication energy cost and a context-dependent optimum exists between in-sensor analytics and communication. It has been shown recently [13], that by tracking this optimum energy-point a IoT Wireless Video Sensor Node can achieve 4.3x improvement in energy efficiency.

## 3.2 Security

We would like to see active development of usable security solutions for this space. These security solutions will span the range in the following dimensions: (i) resource consumption (compute, network communication); (ii) level of security (*e.g.*, does it provide protection against replay attacks? How much of a brute force attack can it tolerate?); (iii) level of user intervention required (does the user need to type in a 6-digit pin, or is only a directional pointing of the device enough?); (iv) use of a trusted third party (does the protocol require intermediation of a trusted third party? This is an active consideration in mobile payment systems where different product offerings keep a lot, little, or no trusted information with the vendors like Google, Apple, or Samsung).

An important unmet need for security solutions is context awareness. One would want not to have to spend precious energy resources on a security protocol (which can often involve expensive network communication) if the environment is benign. For example, if there are several interfering sources of wireless communication, with potentially malicious intent, then a higher level of security posture may be warranted than in a benign environment. One important question is to what extent should the system automatically infer the context and to what extent, this should be input by the user. We should take care that the cure should not become more damaging than the malaise, *i.e.*, inferring context should not become more resource consuming than in the baseline mode.

# 4. ACKNOWLEDGEMENTS

Authors Bagchi and Koo acknowledge the support from the National Science Foundation through the NeTS program (grant numbers CNS-1409506 and CNS-1409589) as well from AT&T through their Virtual University Research Initiative (VURI), which were used to carry out the activities described in this paper. Author Sen acknowledge SRC SLD and NSF CRII programs for their support. Any opinions, findings, and conclusions or recommendations expressed in this material are those of the authors and do not necessarily reflect the views of the sponsors.